\documentclass{ws-p8-50x6-00}

\begin{document}

\title{Event shape studies, \boldmath{$\alpha_s$} and its running from LEP}

\author{U. Flagmeyer}

\address{BU-GH Wuppertal,  
Gau\ss stra\ss e 20, 42097 Wuppertal, Germany\\ 
E-mail: Uwe.Flagmeyer@cern.ch}


\maketitle

\abstracts{
Data from electron positron annihilations at different center of mass energies
are collected by the four LEP experiments. Direct tests 
of hard QCD are performed by measuring and analysing event shape 
observables. 
}

\section{Introduction}
From LEP1 data millions of hadronic events at and around
$M_Z$ are collected and analysed by the four LEP experiments. By selecting
events with prompt and hard photon radiation the energy range below $M_Z$ is
accessible. 
At LEP2 energies the statistics is about three orders of magnitudes smaller
compared to 
LEP1, but still sufficient for measurements of event shape observables. 
The background conditions at LEP2 are much more
various and complicated compared to LEP1. Events with initial state photon
radiation and (semi-) hadronically decaying W- and Z-pairs have to be
discarded. 

From hadronic events infrared and collinear safe event shape observables
and their mean values are deduced. All measurements are in good agreement with
Monte Carlo simulations, none of the LEP collaborations reports about a
significant excess of multi jet events at neither energy. By comparing the
measurements with theoretical predictions from perturbative QCD the strong
coupling $\alpha_s$ and 
its running as well as the QCD structure constants $C_A$ and $C_F$ are
accessible. The theoretical prediction has to be corrected for hadronisation
effects. Two different approaches are followed by  the LEP collaborations. 
Either the transition from colour charged partons into colourless hadrons is
simulated by Monte Carlo generators or power
corrections are applied. Different generators based on
string and cluster fragmentation, all tuned to LEP1 data have been used.
Within power corrections a universal parameter 
$\alpha_0(\mu_I)$ is
introduced, accounting for non perturbative contributions below an infrared
matching scale $\mu_I$. Power corrections can be applied to most of the
event shape observables, to their shapes as well as to their mean values.

By comparing results at different center of mass energies the energy
dependence of event shape observables and of $\alpha_s$ is accessible.

\section{Results}
\subsection{Measurements of $\alpha_s$}
Measurements of $\alpha_s$ are performed by fitting QCD predictions to event
shape observables 
\cite{ALEPH2001-007,OPALPN469,L3Note2670,ALEPH2001-062,Kluth:2001km,MovillaFernandez:2001ed,DELPHI2001-065,DELPHI2001-059,DELPHI2001-062,Abbiendi:2001qn,ALEPH2001-042}. Predictions in fixed 
order (${\cal O}(\alpha_s^2)$ for
3-jet final states and ${\cal O}(\alpha_s^3)$ for 4-jet final states) and in
NLLA are used as well as resummed predictions (R, $\ln$R matching). The
predictions are folded either by a hadronisation correction with Monte Carlo
Models (Pythia, Ariadne, Herwig, Apacic$++$) or by power corrections. The
DELPHI collaboration is performing fits with optimised renormalisation scales,
leeding to much more consistent results.

Using power corrections fits to event shape means leads to
significantly larger results in $\alpha_s$ compared to fit results to event
shape distributions. The expected universality of the parameter $\alpha_0$ is
marginaly fullfilled. 

\begin{figure}[tbhp]
\epsfxsize=8cm 
\begin{center}
\epsfbox{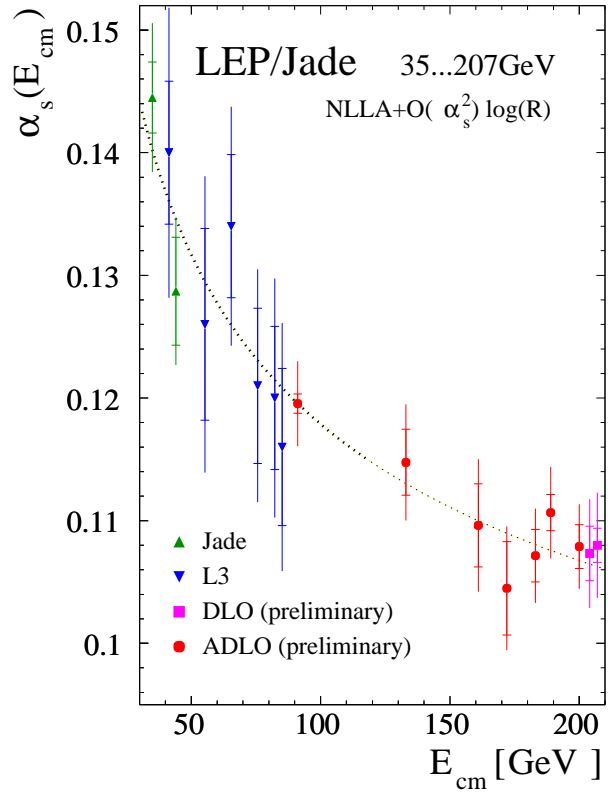} 
\end{center}
\caption{The running of the strong coupling $\alpha_s$
  \label{fig:running_of_alphas}} 
\end{figure}

The LEP QCD working group is working on unifying the analyses to merge the
results. Inputs from the experiments with separated staistical, experimental,
systematic, hadronisation and scale uncertainties allow to combine the single
results by taking correlations into account.
From highest energetic LEP data of the year 2000 
an average value of $\alpha_s(206GeV)=0.1080\pm0.0014\pm0.0043$,
corresponding to $\alpha_s(M_Z)=0.1210\pm0.0018\pm0.0054$ is derived. 
By comparing $\alpha_s$ measurements at
different energies the running of $\alpha_s$ is accesible, being in good
agreement with the QCD expectation (Fig.\ref{fig:running_of_alphas}).

An overall $\alpha_s$ fit to LEP data yields: $\alpha_s(M_Z)=0.1195\pm0.0007\pm0.0048$
with a reasonable $\chi^2/\mathrm{ndf}=31.3/36$. The measurement uncertainties
are dominated by systematics.

\subsection{Measurements of $C_A$ and $C_F$}
By performing QCD fits with $C_A$ and $C_F$ $(n_f)$ as free parameters, the
structure constants of the QCD are accessible
\cite{Kluth:2001km,Abbiendi:2001qn,ALEPH2001-042}. Measurements are performed 
either to mainly 4-jet observables from LEP1 data or to the energy dependence
of the mean values of event shape observables. In the latter case also data 
from lower energetic $e^+e^-$ experiments are included. All measurements
confirm the SU(3) being the gauge symmetry of QCD. The measurement
uncertainties are dominated by systematics.

\subsection{Renormalisation Group Invariant measurements}
The DELPHI collaboration presents an analysis \cite{DELPHI2001-062} with a
renormalisation group invariant approach, where scale and scheme dependencies
cancel out completely. A very surprising result is that rgi fits are able to
describe the energy evolution of event shape means without any need of
hadronisation corrections, implying that power corrections to event shape
means mainly parametrise terms which can in principle be calculated
perturbatively. This approach gives the most precise measure of the
$\beta$-function of QCD. By including data from lower energetic $e^+e^-$
experiments the first coefficient can be determined to be $\beta_0=7.9\pm0.3$,
corresponding to $n_f=4.8\pm0.4$. 

\section{Summary}
Event shape distributions at centre of mass energies from 41GeV to 206GeV are
measured from LEP data and are used for direct tests of hard QCD. The strong
coupling is derived from the data and measured to be
$\alpha_S(M_Z)=0.1195\pm0.0007\pm0.0048$. The running of $\alpha_s$ is in good
agreement with the QCD expectation. The measurement of the structure constants
$C_A$ and $C_F$ underline the SU(3) being the gauge symmetry of QCD. 

Within the analyses of the four LEP collaborations there are still many 
options:
Theoretical predictions are used in fixed order and in NLLA,
different matching schemes are applied. Hadronisation corrections are applied 
by power corrections or by Monte Carlo generators with different tunings.
Scale optimisation, giving a more consistent measure of $\alpha_s$, are 
only slowly accepted by all experiments. 
There is still work to be done to unify the analyses to merge their 
results.

\end{document}